\documentclass{article}
\usepackage{spconf,amsmath,graphicx,hyperref}
\usepackage{enumitem}
\usepackage{multirow}
\usepackage{booktabs}
\usepackage{makecell}
\usepackage{pifont}
\usepackage[table]{xcolor} 
\usepackage{xcolor}
\usepackage{amssymb}

\title{Can Large Audio Language Models Understand Audio Well? \\ Speech, Scene and Events Understanding Benchmark for LALMs}
\name{Han Yin$^{1}$ and Jung-Woo Choi$^{1}$\sthanks{Corresponding author (jwoo@kaist.ac.kr) }}
\address{$^1$School of Electrical Engineering, KAIST, Daejeon, Republic of Korea\\
}

\begin{document}
\ninept
\maketitle
\begin{abstract}
Recently, Large Audio Language Models (LALMs) have progressed rapidly, demonstrating their strong efficacy in universal audio understanding through cross-modal integration.  
To evaluate LALMs' audio understanding performance, researchers have proposed different benchmarks.
However, key aspects for real-world interactions are underexplored in existing benchmarks, i.e., audio signals typically contain both speech and non-speech components, and energy levels of these components can vary significantly across different scenarios.
Moreover, most benchmarks do not consider the joint understanding of speech, scene, and events within the same audio clip.
In this work, we introduce \textbf{SSEU-Bench}, the first versatile audio understanding benchmark that explicitly accounts for energy differences between speech and non-speech audio, with both independent and joint understanding settings for speech, scene, and events.
Furthermore, we demonstrate that some LALMs tend to underperform on certain tasks in a joint understanding setting. To address this issue, we introduce Chain-of-Thought, which effectively improves LALMs' joint audio understanding performance by decomposing complex tasks into simpler reasoning steps\footnote{\url{https://sites.google.com/view/sseu-bench}\label{website}}.
\end{abstract}
\section{Introduction}
\label{sec:intro}
Large Audio Language Models (LALMs) are a class of foundation models that jointly process and understand both audio signals and natural language texts~\cite{survey_2024, uniaudio_2024, eval_2025}. 
Recently, LALMs have made significant progress, leading to various state-of-the-art (SOTA) models that demonstrate competitive performance in different tasks such as meeting transcription~\cite{speakerlm-2025} and human–machine interaction~\cite{LTU_2024,LTU-AS_2023}.

Researchers have proposed various benchmarks to evaluate the audio understanding ability of LALMs.
For example, OpenAQA~\cite{LTU_2024} is a widely used audio understanding dataset that focuses on non-speech audio understanding, covering both closed-ended and open-ended evaluation sets.
Specifically, the closed-ended set is designed for tasks including Audio Captioning (AC), Sound Event Detection (SED), and Audio Tagging (AT), while the open-ended set treats the problem as Audio Question Answering (AQA), where question-answer pairs are generated by Large Language Models (LLMs) and humans from the audio metadata.
Based on OpenAQA, Open-ASQA~\cite{LTU-AS_2023} introduces additional speech clips and speech-related tasks, such as speaker emotion recognition and gender classification.
More recently, AudioBench~\cite{audiobench_2024}, AIR-Bench~\cite{air-bench_2024}, and MMAU~\cite{mmau_2025} have been proposed to expand the evaluation scope by incorporating more audio reasoning tasks, illustrating a growing trend toward evaluating LALMs across various scenarios.

Despite these advances, existing benchmarks do not consider two important characteristics of real-world human-machine interactions.
Specifically, in real-world interactive scenarios, audio signals often contain both foreground user speech and background sound events, with potentially large differences in amplitude. This imbalance can be quantified using the signal-to-noise ratio (SNR), which measures how much the foreground speech dominates or is masked by background sounds.
For example, a quiet home typically has a high SNR, where the user's speech is clearly distinguishable, whereas a metro station may present a low SNR, where background sounds dominate.
The ability of LALMs to understand both speech and non-speech content under varying SNR conditions remains unexplored.
Furthermore, most existing benchmarks rarely perform a joint analysis of \textbf{speech}, \textbf{scene}, and \textbf{events} within the same audio sample, with each typically addressed as a separate task.

\begin{figure*}[t]
\centering
\includegraphics[width=1.45\columnwidth]{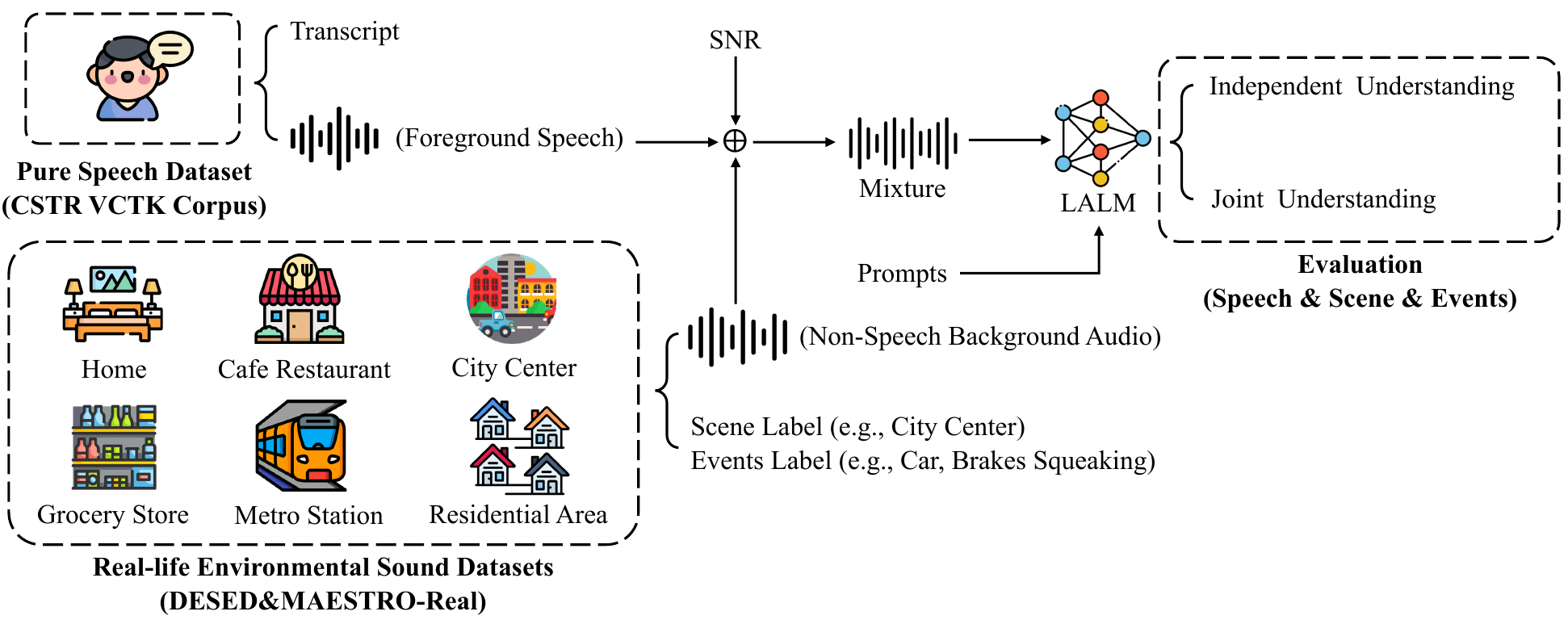}
\caption{Overview of the proposed Speech, Scene and Events Understanding Benchmark (SSEU-Bench).}
\label{fig1:overview}
\end{figure*}

To address these gaps, we propose a new audio understanding benchmark, namely \textbf{Speech, Scene and Events Understanding Benchmark (SSEU-Bench)}.
Specifically, each audio sample in SSEU-Bench is composed of foreground speech and background sound events, mixed at different SNR levels.
We curate realistic background non-speech audio clips from existing SED datasets across 6 scenes (i.e., home, restaurant, city center, grocery store, metro station, and residential area), covering 18 non-speech sound event classes.
Foreground speech is selected from the VCTK corpus~\cite{vctk_2019} and represents clean speech in real-world interactions.
To evaluate the audio understanding capabilities of LALMs, we present three tasks from different understanding perspectives: Automatic Speech Recognition (ASR), Acoustic Scene Classification (ASC), and Audio Tagging (AT). These three tasks enable the LALM to comprehensively interpret audio by addressing three key questions: ``\textit{What is the speaker saying?}'', ``\textit{Where is the speaker (acoustic scene)?}'' and ``\textit{What kinds of events are occurring in the surrounding environment?}''. In addition, we present a Chain-of-Thought (CoT)-guided inference method for LALMs, to improve the joint interpretation of audio information across different granularities. The main contributions of this work are listed as follows:


\begin{itemize}[leftmargin=*,noitemsep]
    \item We present SSEU-Bench. To the best of our knowledge, this is the first audio understanding benchmark to jointly analyze speech, scene, and events within the same audio clip, integrating objective evaluations while explicitly accounting for the energy difference between foreground speech and background sound events. Specifically, we present three tasks, i.e., ASR, ASC and AT, to benchmark the audio understanding performance of LALMs.
    \item Based on SSEU-Bench, we introduce a CoT-guided inference approach to enhance joint understanding of speech, scene, and events for LALMs. This method decomposes audio reasoning into sequential steps before generating the final structured output, improving the overall interpretability and accuracy.
    \item We release all data and code to promote research on joint understanding of speech, scene and events through LALMs.
\end{itemize}

\section{SSEU-Bench}
\subsection{Audio Mixture Creation}


In SSEU-Bench, each audio sample is mixed with a pure speech clip and a non-speech background audio clip with a specific SNR (Fig.~\ref{fig1:overview}).
``SNR'' refers to the strength of the foreground speech relative to the background non-speech audio. 
Specifically, clean speech clips are sourced from VCTK corpus, while non-speech background clips are sampled from DESED\footnote{\url{https://project.inria.fr/desed/}} and MAESTRO-Real~\cite{maestro-real-2023}.

All recordings are converted to a single channel at a sampling rate of 44.1 kHz during mixing.
For inference, recordings are resampled to the sampling rate expected by each model if it differs from 44.1 kHz. 
VCTK provides accurate text transcriptions for each speech clip. 
DESED and MAESTRO-Real offer event labels with timestamps. 
After removing overlapping event categories between the two datasets, we retain 18 distinct sound event classes, such as ``\textit{Alarm Bell Ringing}'' and  ``\textit{Brakes Squeaking}''.
DESED is primarily focused on ``\textit{Home}'' scene, whereas MAESTRO-Real covers five different scenes.
Overall, we sample 378 and 739 non-speech audio clips from the DESED and MAESTRO-Real, respectively. 
From VCTK, we sample 2.71 hours of clean speech data spoken by 104 different speakers. 
In total, we obtain 21.72 hours of audio data, covering seven different SNR conditions.
Detailed scene and event class lists are provided on the benchmark website\textsuperscript{\ref{website}}.

\subsection{Audio Understanding Evaluation}
To evaluate the audio understanding capability of LALMs, we present three different tasks, including ASR, ASC and AT.

\textbf{ASR} aims to evaluate the speech understanding performance of LALMs. 
We input the audio signal along with a text prompt into the LALM to perform the ASR task, where the model is required to transcribe the spoken content accurately. This task is designed to investigate the robustness of LALMs in speech recognition under varying levels of non-speech background audio interference.

\begin{figure}[t]
\centering
\includegraphics[width=1\columnwidth]{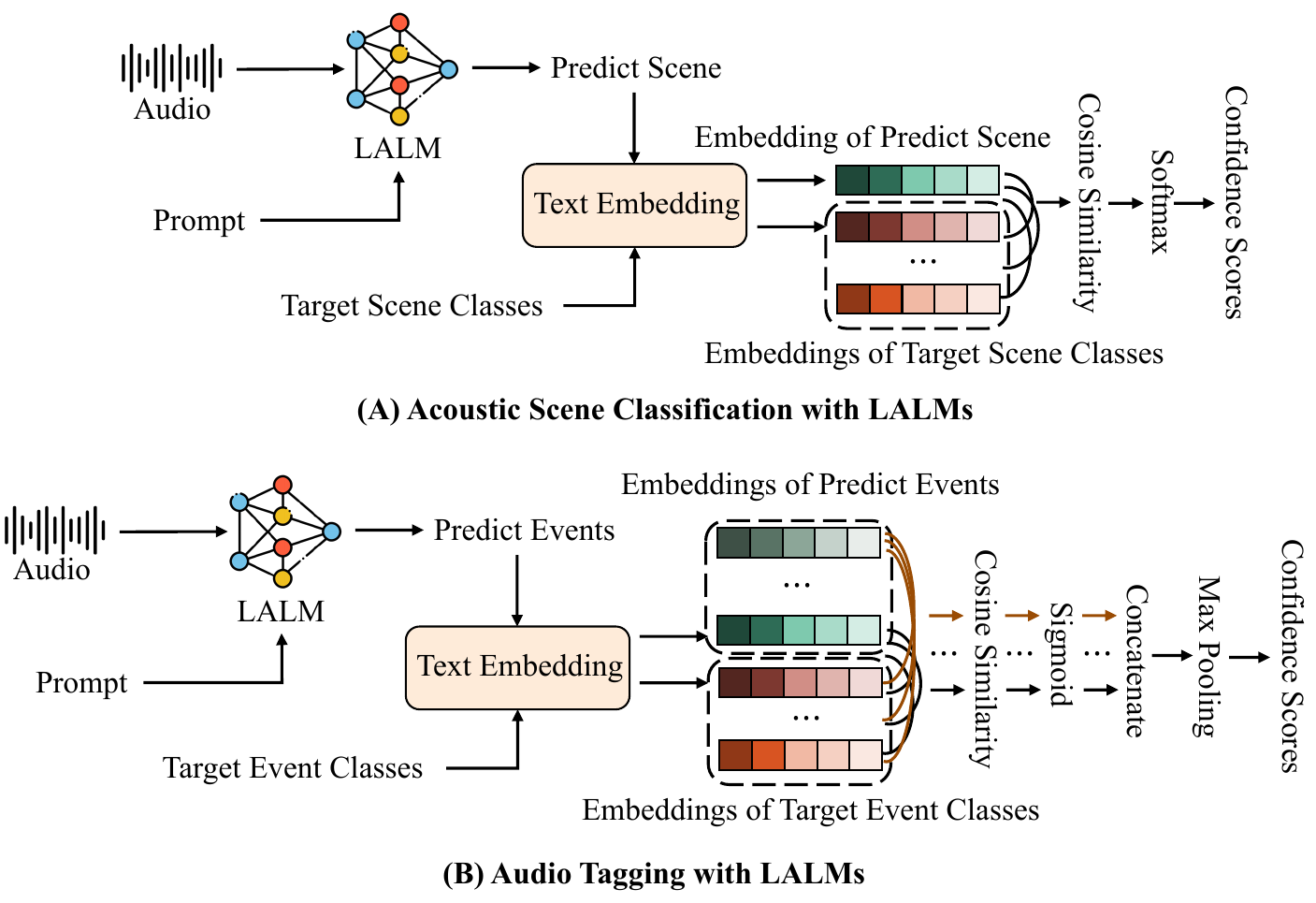}
\caption{Overview of the Scene and Events Understanding Evaluation Methods in SSEU-Bench.}
\label{fig2:eval}
\end{figure}

\textbf{ASC} is designed to analyze LALMs' performance in understanding the acoustic scene. In this task, the LALM is required to identify the environment in which the speaker is located, such as ``\textit{Home}'' or ``\textit{City Center}''. However, existing LALMs are not well-suited for directly generating confidence scores for classification tasks~\cite{LTU_2024,probabilistic_2024}. Therefore, as illustrated in Fig.~\ref{fig2:eval}(A), inspired by previous work~\cite{LTU_2024}, we adopt a text embedding model-based framework to estimate the confidence score for each scene class. Specifically, we use ChatGPT-Text-Embedding-3-Large\footnote{\url{https://platform.openai.com/docs/models/text-embedding-3-large}} as the text embedding model, which is a SOTA embedding model that maps text into high-dimensional vectors, enabling semantic comparison across diverse languages and tasks.

We denote the predicted scene as $\hat{S}$, and define the set of target scene classes as $\mathcal{G}_s =\{ S_1, S_2, \ldots, S_{N_s}\}$, where $N_s=6$ is the number of target scene classes. By expressing the text embedding model as $\theta(\cdot)$ and denoting cosine similarity as $\operatorname{sim}(\cdot)$, the corresponding confidence scores can then be calculated as:
\begin{equation}
    c_i = \frac{e^{\operatorname{sim}(\theta(\hat{S}),\theta(S_i))}}{\sum_{j=1}^{N_s}e^{\operatorname{sim}(\theta(\hat{S}),\theta(S_j))}}~~~(i=1,2,\ldots, N_s)
\end{equation}

\textbf{AT} is used to evaluate the LALM's understanding of sound events in the current environment, providing a more fine-grained assessment compared to ASC. In this task, the LALM is required to identify non-speech sound events present in the audio, such as ``\textit{Cat}'' and ``\textit{Dog}'', among $N_e=18$ target classes. As illustrated in Fig.~\ref{fig2:eval}(B), we still employ a pre-trained text embedding model to compute confidence scores for each target event class.

For the predicted events $\mathcal{P}_e = \{\hat{E}_1, \hat{E}_2, \ldots, \hat{E}_{M_e}\}$, and the target event classes $\mathcal{G}_e = \{E_1, E_2, \ldots, E_{N_e}\}$, we calculate the confidence score matrix $F\in\mathbb{R}^{M_e\times N_e}$ as:
\begin{equation}
    F_{m,n} = \sigma(\operatorname{sim}(\theta(\hat{E}_m),\theta(E_n)))
\end{equation}
where $\sigma(\cdot)$ is the sigmoid function, $m=1,2,\ldots,M_e$ and $n=1,2,\ldots,N_e$. $M_e$ is the number of predicted sound event classes. Then, we apply max pooling to generate the final confidence scores, as formulated in:
\begin{equation}
    p_{n} = \max_{m} F_{m,n}
\end{equation}

To thoroughly investigate the understanding capabilities of LALMs, we propose two evaluation paradigms, namely ``independent understanding'' and ``joint understanding''.
Specifically, for \textbf{independent} understanding, the LALM is required to focus on a single task (i.e., ASR, ASC or AT), while for \textbf{joint} understanding, the LALM is expected to generate predictions for all three tasks.

\subsection{Chain-of-Thought-Guided Inference}

When performing audio understanding, a typical approach is to directly prompt the LALM to generate an answer~\cite{LTU-AS_2023}. 
Recently, CoT has been proposed to enhance complex problem solving by decomposing the task into intermediate steps~\cite{cot_2022}. In this work, we first employ direct prompting to perform joint understanding. Then, we present a CoT-guided inference method that decomposes the joint understanding process into multiple subtasks.
Specifically, before producing the final answer, we guide the LALM through the following reasoning steps: (1) \textbf{Energy \& Onset:} compare the energy between speech and background audio to estimate when speech begins; (2) \textbf{ASR:} transcribe the spoken content accurately; (3) \textbf{Scene candidates:} analyze non-speech audio and generate a ranked list of candidate scenes; (4) \textbf{Event candidates:} detect distinct non-speech sound events and rank candidate events; (5) \textbf{Correction}: cross-validate scenes and events, remove inconsistencies, and select one final scene along with consistent events. Through these steps, the model is expected to refine its understanding and improve prediction accuracy. All prompts used in this work are provided on the website\textsuperscript{\ref{website}}.

\section{Experimental Setups}

\textbf{Large Audio Language Models:} We evaluate four open-source LALMs on SSEU-Bench, including LTU-AS~\cite{LTU-AS_2023}, Qwen2-Audio-Instruct~\cite{qwen2-audio-2024}, Kimi-Audio~\cite{kimi-2025}, and Step-Audio 2 Mini~\cite{step-audio-2}. These models have achieved competitive performance across different audio understanding benchmarks. All four models employ a 7B LLM as the backbone. We present the mean and standard deviation obtained from three runs to assess result variability.

\noindent \textbf{CLAP-based Baselines:} Beyond next-token-prediction-based audio language models (i.e., LALMs), Contrastive Language-Audio Pretraining (CLAP) has also been widely used in downstream audio classification tasks~\cite{clap_2023}.
Therefore, we include various pre-trained CLAP-based models as strong baselines, to compare with LALM-based approaches on ASC and AT. In detail, the audio and class labels are encoded separately; cosine similarities between their embeddings are computed, and the confidence scores are finally obtained using softmax for ASC or sigmoid for AT. We apply the original CLAP~\cite{clap_2023}, and two variants (i.e., LAION-CLAP~\cite{laion_clap_2023} and MGA-CLAP~\cite{mga_clap_2024}), as the CLAP-based strong baselines. 

\noindent \textbf{Metrics:} For the three tasks (i.e., ASR, ASC, and AT), we use Word Error Rate (WER), macro-average Accuracy (mACC), and mean Average Precision (mAP) as the evaluation metrics, respectively, which have been widely applied in previous studies~\cite{speakerlm-2025, panns-2020}. All metrics are reported as percentages in this work.

\noindent \textbf{Upper Bound:} To verify the effectiveness of the proposed evaluation methods and obtain upper-bound metrics, we use the ground truth as predictions and compute corresponding metrics for the three tasks, denoted as WER$_{\textrm{gt}}$, mACC$_{\textrm{gt}}$, and mAP$_{\textrm{gt}}$, respectively.

\noindent \textbf{SNR Selection:} Different acoustic scenes exhibit distinct SNR characteristics. However, since there is currently no publicly available data on real SNR distributions across different scenes, we uniformly assign SNRs ranging from $–$10 dB to 10 dB for each scene. 








\begin{table*}[ht!]
\centering
\caption{The independent understanding performance of different systems on SSEU-Bench. 
``SNR=$+\infty$'' and ``SNR=$-\infty$'' represent ``pure foreground speech'' and ``pure background audio'', respectively.
}
\renewcommand\arraystretch{0.7}{
\setlength{\tabcolsep}{1.7mm}{
\scalebox{1}{
\begin{tabular}{c|ccccccc|c}
\toprule
\multirow{2}{*}{\textbf{System}} & \multicolumn{7}{c|}{\textbf{SNR}} & \multirow{2}{*}{\textbf{Average}} \\
& $+\infty$ & 10 dB & 5 dB & 0 dB & $-$5 dB & $-$10 dB & $-\infty$ & \\
\midrule
\multicolumn{9}{l}{\textbf{Speech Understanding} (WER~$\downarrow$) (\textit{Using the ground-truth transcripts as predictions: WER$_{\textrm{gt}}$ = 0.00\%})} \\
\midrule
LTU-AS~\cite{LTU-AS_2023} & 100.93$\pm$1.2 & 83.34$\pm$3.6 & 81.35$\pm$1.6 & 84.03$\pm$0.7 & 87.24$\pm$1.0 & 98.85$\pm$1.8 & - & 89.29$\pm$0.3 \\
Qwen2-Audio-Instruct~\cite{qwen2-audio-2024} & 4.02$\pm$0.1 & 6.18$\pm$0.1 & 7.86$\pm$0.1 & 11.19$\pm$0.2 & 20.02$\pm$0.2 & 37.71$\pm$0.2 & - & 14.50$\pm$0.1\\
Kimi-Audio~\cite{kimi-2025} & 1.74$\pm$0.1 & 2.43$\pm$0.1 & 2.97$\pm$0.3 & 4.20$\pm$0.2 & 10.67$\pm$0.2 & 30.69$\pm$0.2 & - & \textbf{8.78}$\pm$0.1\\
Step-Audio 2 Mini~\cite{step-audio-2} & 1.91$\pm$0.1 & 10.41$\pm$0.3 & 13.30$\pm$0.8 & 16.48$\pm$1.4 & 25.32$\pm$1.0 & 45.84$\pm$0.8 & - & 18.88$\pm$0.5\\
\midrule
\multicolumn{9}{l}{\textbf{Scene Understanding} (mACC~$\uparrow$) (\textit{Using the ground-truth scene labels as predictions: mACC$_{\textrm{gt}}$ = 100\%})} \\
\midrule
\cellcolor{gray!15} CLAP~\cite{clap_2023} & \cellcolor{gray!15} - & \cellcolor{gray!15} 45.45 & \cellcolor{gray!15} 51.43 & \cellcolor{gray!15} 56.58 & \cellcolor{gray!15} 59.85 & \cellcolor{gray!15} 61.08 & \cellcolor{gray!15} 66.70 & \cellcolor{gray!15} 56.85\\
\cellcolor{gray!15} LAION-CLAP~\cite{laion_clap_2023} & \cellcolor{gray!15} - & \cellcolor{gray!15} 60.77 & \cellcolor{gray!15} 63.73 & \cellcolor{gray!15} 67.19 & \cellcolor{gray!15} 68.10 & \cellcolor{gray!15} 71.04 & \cellcolor{gray!15} 76.61 & \cellcolor{gray!15} \textbf{67.91}\\
\cellcolor{gray!15} MGA-CLAP~\cite{mga_clap_2024} & \cellcolor{gray!15} - & \cellcolor{gray!15} 48.46 & \cellcolor{gray!15} 50.08 & \cellcolor{gray!15} 50.33 & \cellcolor{gray!15} 51.26 & \cellcolor{gray!15} 53.10 & \cellcolor{gray!15} 65.65 & \cellcolor{gray!15} 53.15\\
\midrule
LTU-AS~\cite{LTU-AS_2023} & - & 20.86$\pm$0.2 & 21.09$\pm$0.1 & 21.78$\pm$0.3 & 22.29$\pm$0.1 & 23.34$\pm$0.1 & 32.22$\pm$0.3 & 23.60$\pm$0.1\\
Qwen2-Audio-Instruct~\cite{qwen2-audio-2024}  & - & 29.44$\pm$0.3 & 28.65$\pm$0.2 & 30.67$\pm$0.2 & 32.94$\pm$0.1 & 34.49$\pm$0.2 & 38.67$\pm$0.1 & 32.48$\pm$0.1\\
Kimi-Audio~\cite{kimi-2025} & - & 19.33$\pm$0.4 & 20.56$\pm$1.0 & 21.46$\pm$0.3 & 23.03$\pm$0.2 & 25.68$\pm$0.6 & 38.64$\pm$0.9 & 24.78$\pm$0.2\\
Step-Audio 2 Mini~\cite{step-audio-2} & - & 29.38$\pm$0.3 & 32.11$\pm$0.3 & 35.43$\pm$0.7 & 38.97$\pm$0.6 & 42.78$\pm$0.6 & 50.54$\pm$0.4 & \textbf{38.20}$\pm$0.1\\
\midrule
\multicolumn{9}{l}{\textbf{Events Understanding} (mAP~$\uparrow$) (\textit{Using the ground-truth event labels as predictions: mAP$_{\textrm{gt}}$ = 100\%})} \\
\midrule
\cellcolor{gray!15} CLAP~\cite{clap_2023} &\cellcolor{gray!15} - & \cellcolor{gray!15} 55.19 & \cellcolor{gray!15} 57.14 & \cellcolor{gray!15} 57.86 & \cellcolor{gray!15} 57.10 & \cellcolor{gray!15} 57.28 & \cellcolor{gray!15} 63.11 & \cellcolor{gray!15} 57.95\\
\cellcolor{gray!15} LAION-CLAP~\cite{laion_clap_2023} & \cellcolor{gray!15} - & \cellcolor{gray!15} 60.33 & \cellcolor{gray!15} 61.35 & \cellcolor{gray!15} 61.80 & \cellcolor{gray!15} 62.13 & \cellcolor{gray!15} 62.41 & \cellcolor{gray!15} 65.20 & \cellcolor{gray!15} 62.20\\
\cellcolor{gray!15} MGA-CLAP~\cite{mga_clap_2024} & \cellcolor{gray!15} - & \cellcolor{gray!15} 62.09 & \cellcolor{gray!15} 62.56 & \cellcolor{gray!15} 62.87 & \cellcolor{gray!15} 63.12 & \cellcolor{gray!15} 63.49 & \cellcolor{gray!15} 71.64 & \cellcolor{gray!15} \textbf{64.29} \\
\midrule
LTU-AS~\cite{LTU-AS_2023} & - & 23.09$\pm$0.3 & 23.42$\pm$0.4 & 25.05$\pm$0.3 & 26.12$\pm$0.2 & 27.67$\pm$0.4 & 36.95$\pm$0.1 & 27.05$\pm$0.1 \\
Qwen2-Audio-Instruct~\cite{qwen2-audio-2024} & - & 30.78$\pm$0.2 & 30.91$\pm$0.4 & 33.59$\pm$0.3 & 35.59$\pm$0.6 & 36.27$\pm$0.8 & 40.09$\pm$0.7 & 34.54$\pm$0.2\\
Kimi-Audio~\cite{kimi-2025} & - & 21.65$\pm$0.1 & 22.03$\pm$0.4 & 23.20$\pm$0.5 & 25.67$\pm$0.6 & 28.46$\pm$0.3 & 38.06$\pm$0.5 & 26.51$\pm$0.2\\
Step-Audio 2 Mini~\cite{step-audio-2} & - & 33.45$\pm$0.2 & 34.22$\pm$1.1 & 36.02$\pm$0.3 & 38.95$\pm$0.9 & 41.03$\pm$0.6 & 45.00$\pm$0.7 & \textbf{38.11}$\pm$0.3\\
\bottomrule
\end{tabular}}
}
}
\label{tab:main}
\end{table*}

\section{Results and Discussions}

\textbf{Independent Understanding:} We first present the independent understanding performance of LALMs and CLAP-based models in Table~\ref{tab:main}.
For speech understanding, LTU-AS exhibits an extremely high WER (i.e., over 80\%). This is because LTU-AS uses a pre-trained ASR model (i.e., Whisper~\cite{whisper_2023}) to transcribe the speech into text, which is then injected into the LLM backbone. When the spoken text is not provided, its ASR performance drops dramatically~\cite{LTU-AS_2023}, with the WER on LibriSpeech~\cite{librispeech_2015} increasing from 4.9\% to 97.2\%.
In addition, the WER on clean speech is higher than that on other conditions, which is because the ASR training data for LTU-AS is derived from AudioSet~\cite{audioset_2017}, where most samples are mixtures of environmental sounds and speech instead of clean speech.
Both Qwen2-Audio-Instruct and Step-Audio 2 Mini achieve competitive performance on clean speech, but their ASR results are severely affected by background audio interference. Kimi-Audio demonstrates the most robust ASR performance, showing minimal sensitivity to non-speech signals and achieving the lowest average WER of 8.78\%.

For scene understanding, CLAP-based methods outperform LALM-based counterparts, as CLAP naturally aligns audio and text in a shared embedding space, which is beneficial for zero-shot classification and retrieval tasks~\cite{clap_2023}. 
However, although CLAP-based methods can generalize to unseen labels, they rely on a pre-defined label set during inference. 
For LALMs, we do not require them to generate pre-defined labels. 
Consequently, the outputs of LALMs may not exactly match the scene labels, and therefore a pre-trained text embedding model is applied to estimate the similarity between the predicted and target scenes. 
Step-Audio 2 Mini achieves the best ASC result, with an average mACC of 38.2\%.
A similar trend is observed in events understanding, where CLAP-based models outperform LALMs, and Step-Audio 2 Mini achieves the SOTA performance, with an average mAP of 38.11\%.

\begin{table}[htbp]
\centering
\vspace{-6pt}
\caption{The independent and joint understanding performance of different LALMs on SSEU-Bench.}
\renewcommand\arraystretch{0.8}{
\setlength{\tabcolsep}{0.3mm}{
\scalebox{1}{
\begin{tabular}{c|c|c|c|c}
\toprule
System & Joint & WER~$\downarrow$ & mACC~$\uparrow$ & mAP~$\uparrow$\\
\midrule
\multirow{2}{*}{Qwen2-Audio-Instruct~\cite{qwen2-audio-2024}} & \ding{55} & \textbf{16.59}$\pm$0.1 & \textbf{31.24}$\pm$0.1 & \textbf{33.42}$\pm$0.1\\
& \ding{51} & 22.16$\pm$0.4 & 21.86$\pm$0.1 & 16.52$\pm$1.1\\
\midrule
\multirow{2}{*}{Kimi-Audio~\cite{kimi-2025}} & \ding{55} & \textbf{10.19}$\pm$0.1 & 22.01$\pm$0.2 & 24.20$\pm$0.3\\
& \ding{51} & 17.84$\pm$0.6 & \textbf{22.80}$\pm$0.2 & \textbf{26.91}$\pm$0.4\\
\midrule
\multirow{2}{*}{Step-Audio 2 Mini~\cite{step-audio-2}} & \ding{55} &  22.27$\pm$0.5 & \textbf{35.73}$\pm$0.3 & \textbf{36.73}$\pm$0.5\\
& \ding{51} & \textbf{19.01}$\pm$0.2 & 24.39$\pm$0.4 & 30.22$\pm$0.2\\
\bottomrule

\end{tabular}}
}
}
\label{tab:ab_1}
\end{table}

\noindent \textbf{Impact of Joint Understanding:} 
In Table~\ref{tab:ab_1}, we report the independent and joint audio understanding performance of Qwen2-Audio-Instruct, Kimi-Audio, and Step-Audio 2 Mini, where the metrics are averaged results across five different SNR conditions (i.e., 10~dB, 5~dB,  0~dB,  $-$5~dB and $-$10~dB).
We exclude LTU-AS from the comparison, as we observe that when performing joint understanding, LTU-AS fails to generate expected answers for all three tasks as required by the text prompt. Results show that the impact of joint understanding varies across different LALMs.

Specifically, for Qwen2-Audio-Instruct, the performance under joint understanding drops sharply, suggesting that the model is less effective at performing joint speech, scene and events understanding. 
In addition, we observe that Qwen2-Audio-Instruct occasionally produces multiple timestamps for sound events, which can be attributed to the influence of the SED task during pre-training. For Kimi-Audio and Step-Audio 2 Mini, when performing joint understanding, the models tend to concentrate on certain tasks, leading to performance degradation on the remaining tasks.

\begin{figure}[t]
\centering
\includegraphics[width=1.0\columnwidth]{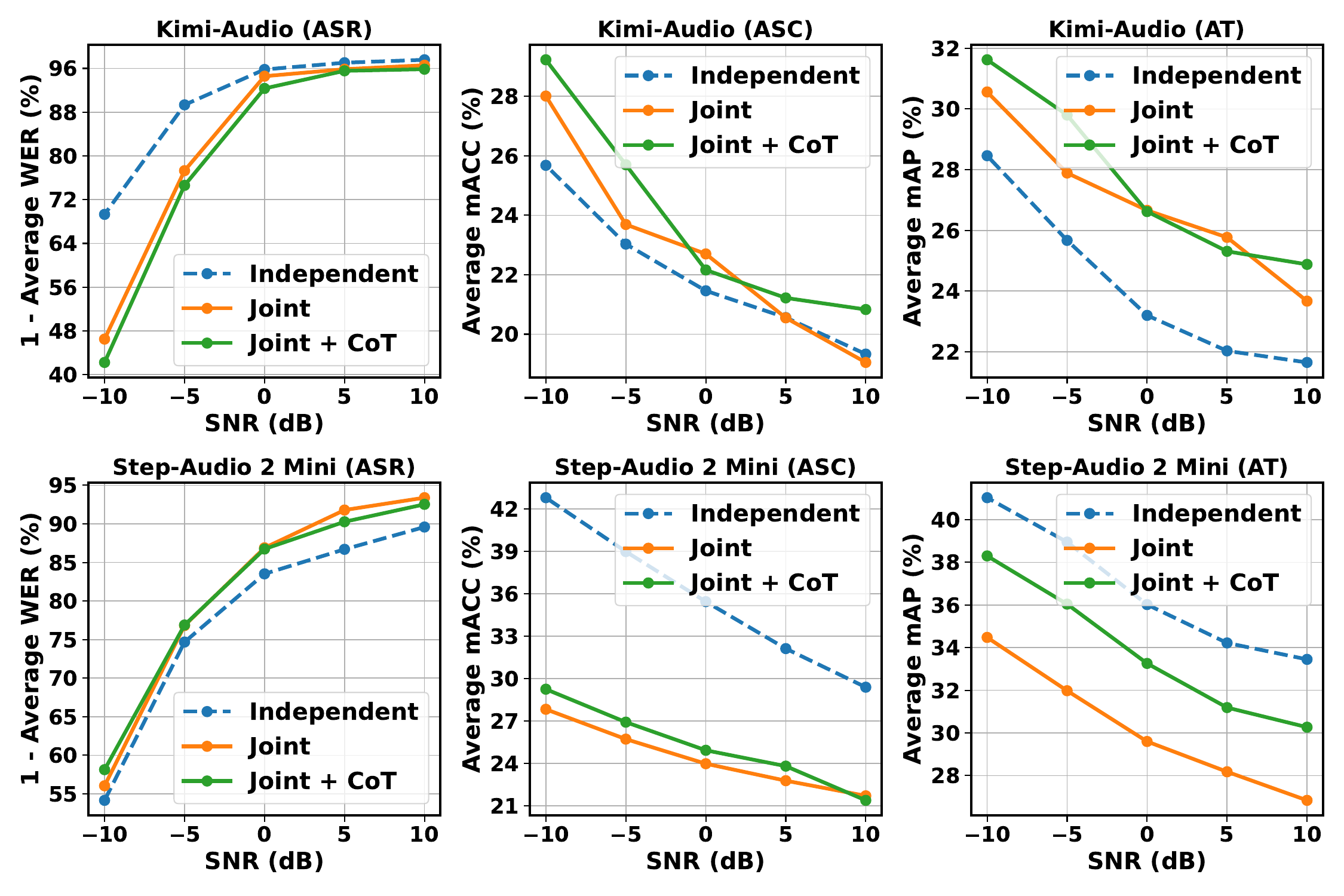}
\caption{Performance of Step-Audio 2 Mini with different understanding strategies on SSEU-Bench.}
\label{fig3:eval}
\end{figure}

\noindent \textbf{Impact of CoT:} We further conduct an in-depth analysis of Step-Audio 2 Mini and Kimi-Audio in Fig.~\ref{fig3:eval}, exploring how CoT affects joint understanding across different conditions. 
We exclude Qwen2-Audio-Instruct here, as the model exhibits unstable behavior under CoT guidance, frequently failing to produce the expected outputs.

Results show that CoT is beneficial for improving the overall joint audio understanding performance.
In detail, for Kimi-Audio, CoT leads to relatively stable ASR performance, while both ASC and AT exhibit improvements under certain SNR conditions (e.g, 10 dB, –5 dB, and –10 dB). 
This gain can be attributed to the last step introduced in CoT, where the model is guided to leverage the correlation between scenes and events to correct the corresponding results. 
Such correction is particularly beneficial for ASC and AT, whereas ASR, as the first step of reasoning without subsequent adjustment, shows limited room for improvement. Similarly, Step-Audio 2 Mini tends to focus on a single task (i.e., ASR) under the joint understanding setting, leading to a decline in performance on the other tasks (i.e., ASC and AT). When CoT guidance is applied to the LALM to perform joint understanding step by step, the performance of ASC and AT is improved. Notably, for AT, the average mAP increases by approximately 4\% across all SNR conditions. 
Detailed inference demos are presented on the benchmark website\textsuperscript{\ref{website}}.

\section{Conclusions}

In this paper, we present SSEU-Bench, the first audio understanding benchmark for LALMs that accounts for speech and non-speech audio with energy differences, reflecting realistic interactive scenarios.
Results show that LALMs can achieve promising results on single-granularity (i.e., speech, scene or events) audio understanding tasks, but there is still a significant gap compared with CLAP-based models.
In addition, for joint understanding, some LALMs tend to exhibit degraded performance on certain tasks. 
By guiding the inference process with CoT, the joint audio understanding performance can be enhanced.  
These findings underline the promise of LALMs in audio understanding, while also revealing open challenges in unified perception of speech, acoustic scene and sound events. 


\newpage

\section{Acknowledgments}

This work was supported by the National Research Foundation of Korea (NRF) grant (No. RS-2024-00337945); and the BK21 FOUR program through the NRF grant funded by the Ministry of Education of Korea government (MOE).
\bibliographystyle{IEEEbib}
\bibliography{refs}

@article{eval_2025,
  title={Towards holistic evaluation of large audio-language models: A comprehensive survey},
  author={Yang, Chih-Kai and Ho, Neo S and Lee, Hung-yi},
  journal={arXiv preprint arXiv:2505.15957},
  year={2025}
}

@article{survey_2024,
  title={A survey on multimodal large language models},
  author={Yin, Shukang and Fu, Chaoyou and Zhao, Sirui and Li, Ke and Sun, Xing and Xu, Tong and Chen, Enhong},
  journal={National Science Review},
  volume={11},
  number={12},
  pages={nwae403},
  year={2024},
  publisher={Oxford University Press}
}

@inproceedings{audioset_2017,
  title={{Audioset: An ontology and human-labeled dataset for audio events}},
  author={Gemmeke, Jort F and Ellis, Daniel PW and Freedman, Dylan and Jansen, Aren and Lawrence, Wade and Moore, R Channing and Plakal, Manoj and Ritter, Marvin},
  booktitle={International Conference on Acoustics, Speech and Signal Processing (ICASSP)},
  pages={776--780},
  year={2017},
  organization={IEEE}
}

@misc{vctk_2019,
  author       = {Yamagishi, Junichi and Veaux, Christophe and MacDonald, Kirsten},
  title        = {{CSTR VCTK Corpus: English Multi-speaker Corpus for CSTR Voice Cloning Toolkit (version 0.92) [sound]}},
  year         = {2019},
  howpublished = {University of Edinburgh. The Centre for Speech Technology Research (CSTR)},
  note         = {\url{https://doi.org/10.7488/ds/2645}}
}

@inproceedings{whisper_2023,
  title={Robust speech recognition via large-scale weak supervision},
  author={Radford, Alec and Kim, Jong Wook and Xu, Tao and Brockman, Greg and McLeavey, Christine and Sutskever, Ilya},
  booktitle={International Conference on Machine Learning (ICML)},
  pages={28492--28518},
  year={2023},
  organization={PMLR}
}

@inproceedings{librispeech_2015,
  title={Librispeech: an asr corpus based on public domain audio books},
  author={Panayotov, Vassil and Chen, Guoguo and Povey, Daniel and Khudanpur, Sanjeev},
  booktitle={International Conference on Acoustics, Speech and Signal Processing (ICASSP)},
  pages={5206--5210},
  year={2015},
  organization={IEEE}
}

@article{panns-2020,
  title={{PANNs: Large-scale pretrained audio neural networks for audio pattern recognition}},
  author={Kong, Qiuqiang and Cao, Yin and Iqbal, Turab and Wang, Yuxuan and Wang, Wenwu and Plumbley, Mark D},
  journal={IEEE/ACM Transactions on Audio, Speech, and Language Processing (TASLP)},
  volume={28},
  pages={2880--2894},
  year={2020},
  publisher={IEEE}
}

@article{speakerlm-2025,
  title={{SpeakerLM}: End-to-End Versatile Speaker Diarization and Recognition with Multimodal Large Language Models},
  author={Yin, Han and Chen, Yafeng and Deng, Chong and Cheng, Luyao and Wang, Hui and Tan, Chao-Hong and Chen, Qian and Wang, Wen and Li, Xiangang},
  journal={arXiv preprint arXiv:2508.06372},
  year={2025}
}

@article{step-audio-2,
  title={{Step-Audio 2 Technical Report}},
  author={Wu, Boyong and Yan, Chao and Hu, Chen and Yi, Cheng and Feng, Chengli and Tian, Fei and Shen, Feiyu and Yu, Gang and Zhang, Haoyang and Li, Jingbei and others},
  journal={arXiv preprint arXiv:2507.16632},
  year={2025}
}

@article{kimi-2025,
  title={{Kimi-Audio Technical Report}},
  author={Ding, Ding and Ju, Zeqian and Leng, Yichong and Liu, Songxiang and Liu, Tong and Shang, Zeyu and Shen, Kai and Song, Wei and Tan, Xu and Tang, Heyi and others},
  journal={arXiv preprint arXiv:2504.18425},
  year={2025}
}

@article{qwen2-audio-2024,
  title={{Qwen2-Audio Technical Report}},
  author={Chu, Yunfei and Xu, Jin and Yang, Qian and Wei, Haojie and Wei, Xipin and Guo, Zhifang and Leng, Yichong and Lv, Yuanjun and He, Jinzheng and Lin, Junyang and others},
  journal={arXiv preprint arXiv:2407.10759},
  year={2024}
}

@article{cot_2022,
  title={Chain-of-thought prompting elicits reasoning in large language models},
  author={Wei, Jason and Wang, Xuezhi and Schuurmans, Dale and Bosma, Maarten and Xia, Fei and Chi, Ed and Le, Quoc V and Zhou, Denny and others},
  journal={Advances in Neural Information Processing Systems (NIPS)},
  volume={35},
  pages={24824--24837},
  year={2022}
}

@article{maestro-real-2023,
  title={Strong labeling of sound events using crowdsourced weak labels and annotator competence estimation},
  author={Mart{\'\i}n-Morat{\'o}, Irene and Mesaros, Annamaria},
  journal={IEEE/ACM Transactions on Audio, Speech, and Language Processing (TASLP)},
  volume={31},
  pages={902--914},
  year={2023},
  publisher={IEEE}
}

@article{probabilistic_2024,
  title={Probabilistic medical predictions of large language models},
  author={Gu, Bowen and Desai, Rishi J and Lin, Kueiyu Joshua and Yang, Jie},
  journal={npj Digital Medicine},
  volume={7},
  number={1},
  pages={367},
  year={2024},
  publisher={Nature Publishing Group UK London}
}

@inproceedings{LTU_2024,
  title={{Listen, Think, and Understand}},
  author={Gong, Yuan and Luo, Hongyin and Liu, Alexander H and Karlinsky, Leonid and Glass, James},
  booktitle={International Conference on Learning Representations (ICLR)},
  year={2024}
}

@inproceedings{LTU-AS_2023,
  title={Joint audio and speech understanding},
  author={Gong, Yuan and Liu, Alexander H and Luo, Hongyin and Karlinsky, Leonid and Glass, James},
  booktitle={Automatic Speech Recognition and Understanding Workshop (ASRU Workshop)},
  pages={1--8},
  year={2023},
  organization={IEEE}
}

@article{air-bench_2024,
  title={{Air-bench: Benchmarking large audio-language models via generative comprehension}},
  author={Yang, Qian and Xu, Jin and Liu, Wenrui and Chu, Yunfei and Jiang, Ziyue and Zhou, Xiaohuan and Leng, Yichong and Lv, Yuanjun and Zhao, Zhou and Zhou, Chang and others},
  journal={arXiv preprint:2402.07729},
  year={2024}
}

@article{audiobench_2024,
  title={Audiobench: A universal benchmark for audio large language models},
  author={Wang, Bin and Zou, Xunlong and Lin, Geyu and Sun, Shuo and Liu, Zhuohan and Zhang, Wenyu and Liu, Zhengyuan and Aw, AiTi and Chen, Nancy F},
  journal={arXiv preprint arxiv:2406.16020},
  year={2024}
}

@inproceedings{mmau_2025,
  title={{MMAU: A} Massive Multi-Task Audio Understanding and Reasoning Benchmark},
  author={Sakshi, S and Tyagi, Utkarsh and Kumar, Sonal and Seth, Ashish and Selvakumar, Ramaneswaran and Nieto, Oriol and Duraiswami, Ramani and Ghosh, Sreyan and Manocha, Dinesh},
  booktitle={International Conference on Learning Representations (ICLR)},
  year={2025}
}

@inproceedings{clap_2023,
  title={{CLAP: L}earning Audio Concepts From Natural Language Supervision},
  author={Elizalde, Benjamin and Deshmukh, Soham and Al Ismail, Mahmoud and Wang, Huaming},
  booktitle={International Conference on Acoustics, Speech and Signal Processing (ICASSP)},
  pages={1--5},
  year={2023},
  organization={IEEE}
}

@inproceedings{laion_clap_2023,
  title={Large-scale contrastive language-audio pretraining with feature fusion and keyword-to-caption augmentation},
  author={Wu, Yusong and Chen, Ke and Zhang, Tianyu and Hui, Yuchen and Berg-Kirkpatrick, Taylor and Dubnov, Shlomo},
  booktitle={International Conference on Acoustics, Speech and Signal Processing (ICASSP)},
  pages={1--5},
  year={2023},
  organization={IEEE}
}

@inproceedings{mga_clap_2024,
  title={Advancing multi-grained alignment for contrastive language-audio pre-training},
  author={Li, Yiming and Guo, Zhifang and Wang, Xiangdong and Liu, Hong},
  booktitle={ACM International Conference on Multimedia (ACM MM)},
  pages={7356--7365},
  year={2024}
}

@inproceedings{uniaudio_2024,
  title={Uniaudio: Towards universal audio generation with large language models},
  author={Yang, Dongchao and Tian, Jinchuan and Tan, Xu and Huang, Rongjie and Liu, Songxiang and Guo, Haohan and Chang, Xuankai and Shi, Jiatong and Bian, Jiang and Zhao, Zhou and others},
  booktitle={International Conference on Machine Learning (ICML)},
  year={2024}
}

\end{document}